%% file: main.tex
\documentclass[11pt]{article}
\usepackage[margin=1in]{geometry}
\usepackage{amsmath,amssymb,amsthm}
\usepackage{booktabs}
\usepackage{float}
\usepackage{graphicx}
\usepackage{xcolor}
\usepackage[numbers,sort&compress]{natbib}
\usepackage[colorlinks=true,allcolors=blue]{hyperref}
\newtheorem{proposition}{Proposition}
\newtheorem{corollary}{Corollary}
\theoremstyle{definition}
\newtheorem{definition}{Definition}

\newcommand{\byte}{\mathbf{b}}
\newcommand{\green}{\mathsf{g}}
\newcommand{\AR}{\mathrm{AR}}
\newcommand{\Can}{\mathcal{C}}
\newcommand{\tiwm}{TI-WM}
\input{artifacts/paper_campaign_20261004/numbers.tex}

\title{Tokenization-Invariant Watermarking:\\
A Byte-Local Method for LLM Text}
\author{Tung-Ling Li\\Independent Researcher\\\texttt{tony51307@gmail.com}}
\date{October 2026}

\begin{document}
\maketitle

\begin{abstract}
We introduce Tokenization-Invariant Watermarking (\tiwm{}), a method that
embeds a watermark during token generation but keys and scores it on
canonicalized UTF-8 bytes. The detector does not invoke the tokenizer.
Canonicalizer-preserving edits leave its score unchanged; other localized
canonical-byte edits affect a bounded number of scoring windows. Across three
fixed keys and 128 prompts per key and model, a deterministic white-box attack
at a 5\% word-edit budget leaves mean attacked-AUC \QwenByteAUC{} on Qwen3-8B
and \GemmaByteAUC{} on Gemma 4 E4B, compared with \QwenUnigramAUC{} and
\GemmaUnigramAUC{} for Unigram with detection-side deletion. These operating
points are not jointly matched on quality and clean detectability. Bias sweeps
show model-dependent distortion costs, and shared-rule controls support a
locality interpretation without establishing a unique advantage of bytes.
Visible insertion outside the canonicalizer's deletion set defeats the default
design. Alphabetic filtering closes that channel but does not resolve source
shift or calibrated detection power. \tiwm{} thus provides a concrete
construction with local scoring guarantees and measured attack-ranking gains,
not a general robustness or low-false-positive-rate deployment guarantee.
\end{abstract}

\section{Introduction}
LLM watermarking embeds a statistical signal in generated text so that machine
authorship can later be detected. It is being deployed at scale (e.g.,
SynthID-Text in Gemini~\citep{dathathri2024synthid}). Yet the dominant designs
key the signal on \emph{tokens}: KGW~\citep{kirchenbauer2023watermark} and
SynthID condition on the preceding token(s); Unigram~\citep{zhao2023provable}
uses a fixed token green-list; SIR~\citep{liu2024semantic} keys on semantic token
embeddings. Because detection re-tokenizes the suspect text, a \emph{single
character edit}, such as an invisible zero-width space, a homoglyph, or a typo, can
re-segment the BPE stream and scramble the signal. \citet{zhang2026character}
formalize this as \emph{attack range} and show that character-level attacks reach
more of the signal than token-level ones, that a genetic-algorithm (GA) removal
attack is highly effective, and, critically, that every fixed \emph{detection-side}
defense (spell-check, OCR, Unicode-normalize, delete) is defeated by an adaptive
compound attack. They leave open the design of tokenization-robust \emph{schemes}.
Recent benchmark work reaches the same broad conclusion from a systems angle:
watermark performance is highly attack- and operation-dependent, and robustness
claims need adaptive, cross-scheme evaluation rather than a single perturbation
or detector threshold~\citep{liang2025watermark,qin2026entrobench}.

\paragraph{Our approach.} We introduce Tokenization-Invariant Watermarking
(\tiwm{}), which moves the watermark's keying and scoring units from tokens to
canonicalized bytes. Generation still samples tokens: a logit bias favors
candidate tokens whose canonical bytes are predominantly green under the
byte-local key. Detection applies the same canonicalization and scores the byte
sequence directly, without recovering the generating model's token boundaries.

Two properties motivate the construction. Canonicalization gives exact score
invariance to edits that preserve the canonical string. Byte-local keying
bounds the count loss from other localized edits, independently of how those
edits change tokenization. Neither property guarantees that enough signal was
embedded or that a threshold controls false positives on a new text source.
Bytes within a sampled token remain dependent; more scored positions are not
an equal increase in independent evidence.

\paragraph{Contributions.}
\begin{enumerate}
\item The \tiwm{} construction: generation-time logit bias with canonical-byte
keying and tokenizer-free detection (\S\ref{sec:method}).
\item An aligned attack-range analysis proving exact canonicalizer invariance
and bounding the scoring impact of localized canonical-byte edits, with
numerical substitution radii (\S\ref{sec:theory}).
\item A deterministic adaptive comparison against KGW, Unigram, and their
detection-side defenses on two model families, with 128 prompts per key, three
fixed keys, bias sweeps, and length scaling. Shared-rule controls test the
locality interpretation without assuming a monotone granularity advantage.
\item An explicit account of the construction's limits under visible insertion,
source shift, and low-FPR detection. Separate follow-up designs and unsuccessful
refinements are reported in the appendix rather than treated as the proposed
method.
\end{enumerate}

\section{Background and threat model}
\paragraph{Green-list watermarks.} KGW partitions the vocabulary into ``green''
and ``red'' tokens using a hash of the previous $h$ tokens and adds a bias
$\delta$ to green logits; detection counts green tokens and applies a one-sided
$z$-test. Unigram fixes a single context-free green-list, trading context secrecy
for robustness. SynthID-Text uses tournament sampling over per-token $g$-values.
SIR maps the semantic embedding of the context to watermark logits for
paraphrase robustness. All operate at the token level.
Recent systems such as WaterPool improve the imperceptibility/efficacy/robustness
trade-off but still inhabit this token-centered design family~\citep{huang2025waterpool}.

\paragraph{Tokenization fragility.} A BPE tokenizer merges frequent byte
sequences; inserting or altering one character can split a token into several
sub-tokens, changing the token identities the detector re-derives. Thus a
character edit affects a wider ``attack range'' than a token edit
\citep{zhang2026character}.

\paragraph{Threat model.} Our primary claim concerns \emph{removal/evasion}: an
adversary with an editing-rate budget and query access to the detector perturbs
watermarked text to evade detection (the setting of \citet{zhang2026character}).
We evaluate a white-box GA that directly optimizes against the detector score.
\emph{Spoofing} (forging the watermark onto unwatermarked text)
is a separate axis; we analyze it in \S\ref{sec:spoof} and test a case-swap
mitigation.

\section{Related work}
\begin{table}[H]\centering\small
\caption{Summary of nearby watermarking research lines.}\label{tab:rw}
\begin{tabular}{lll}
\toprule
line & signal & main defense target\\
\midrule
KGW/SynthID & token & statistical detectability\\
Unigram & token & token-edit robustness\\
WaterPool & token & quality/efficacy trade-off\\
SIR/PASA & semantic & paraphrase\\
rollback repair & token & tokenization mismatch\\
\bottomrule
\end{tabular}
\end{table}

\paragraph{Token-level statistical watermarks.}
KGW~\citep{kirchenbauer2023watermark}, Unigram~\citep{zhao2023provable},
SynthID-Text~\citep{dathathri2024synthid}, and WaterPool~\citep{huang2025waterpool}
all bias or score token choices, differing in context structure, sampling rule,
and how they navigate the imperceptibility/efficacy/robustness trade-off.
EXP-edit aligns a pseudorandom key to a candidate sequence using edit distance,
providing a distinct route to insertion, deletion, and substitution robustness
at higher detection cost~\citep{kuditipudi2024robust}. Three Bricks develops
statistical tests with guarantees at very low false-positive rates and
emphasizes repeated-context handling and calibrated detection
~\citep{fernandez2023three}.

\paragraph{Semantic watermarks.}
SIR~\citep{liu2024semantic} and PASA~\citep{ai2026pasa} move toward
semantic signals to survive meaning-preserving rewrites. They target changes in
meaning and phrasing rather than tokenization mismatch from character-level
surface perturbations.

\paragraph{Character attacks.}
\citet{zhang2026character} give the character-level attack-range analysis and
the adaptive ``adversarial dilemma'' for detection-side defenses. Tokenization
rollback/key-restoration approaches address mismatch after generation by trying
to recover the original tokenization~\citep{yan2025tokenization}. Broad
evaluations such as Watermark under Fire and
EntroBench~\citep{liang2025watermark,qin2026entrobench} study robustness across
attacks, operating points, and watermark designs. MARKMyWORDS emphasizes joint evaluation of quality,
detection length, and tamper resistance~\citep{piet2025markmywords}; MarkSec
further shows that attack rankings can change once text quality and adversary
capabilities are constrained~\citep{li2026marksec}.

\paragraph{Detection calibration and edited text.}
Watermark detection is a hypothesis-testing problem: pivotal statistics give
principled control of false-positive rates and a framework for analyzing power
~\citep{li2025statistical}. Robust detection under human edits has also been
studied using mixture models and truncated goodness-of-fit tests
~\citep{li2025humanedits}; earlier reliability experiments considered human
rewrites, machine paraphrases, and watermarked spans mixed into longer documents
~\citep{kirchenbauer2024reliability}. These methods address different
statistics and edit models from ours. They motivate our human-text calibration
and repetition audits, while our focus is on character-level evasion and
cross-source null failures rather than a new calibrated detector. A recent
deployment-oriented systematization likewise emphasizes threat assumptions,
false-positive control, and distribution shift as distinct design dimensions
~\citep{phan2026survey}.

\paragraph{Spoofing and impossibility.}
Watermark stealing shows that green-list structure can leak from observed
watermarked text~\citep{jovanovic2024watermark}. DITTO demonstrates a distinct
distillation-based spoofing route~\citep{an2026ditto}. Stronger impossibility
results rule out universal robustness under quality-preserving
rewrites~\citep{zhang2024watermarks}.

\section{Method: Tokenization-Invariant Watermarking (\tiwm{})}\label{sec:method}
Let $\Can:\Sigma^\ast\to\Sigma^\ast$ delete a fixed set $D$ of characters
(whitespace and zero-width/format characters), optionally case-folding first
(\S\ref{sec:spoof}). Let $\byte(x)=\mathrm{utf8}(\Can(x))\in\{0,\dots,255\}^n$ be
the canonical byte string. Fix a key $k$, green fraction $\gamma$, and byte-context
order $L\ge 0$. For byte position $t\in[L,n)$ with context $c_t=\byte(x)[t{-}L{:}t]$,
a keyed hash $\mathrm{SHA256}(k\,\|\,c_t)$ seeds a green set
$G(k,c_t)\subset\{0,\dots,255\}$ of size $\lfloor\gamma\cdot256\rceil$. The
byte indicator is $\green_t(x)=\mathbf 1[\byte(x)[t]\in G(k,c_t)]$.

\paragraph{Generation.} At each step we add $\delta\cdot\mathbf 1[\text{greenness}(w)>\tfrac12]$
to the logit of candidate token $w$, where greenness is the fraction of $w$'s
canonical bytes that are green under the sliding context (each byte scored under
the green set seeded by the bytes immediately preceding it, exactly as detection
will score it).
This majority-green rule is the generation rule used for the primary
\tiwm{} comparison. Alternative count rewards and frequency-balanced green
sets are follow-up ablations, described in Appendix~\ref{app:refinements}.

\paragraph{Detection.} Compute $S=\sum_{t=L}^{n-1}\green_t(x)$ and
$z=(S-pT)/\sqrt{Tp(1-p)}$, $T=n-L$; declare watermarked if $z>\tau$. Over a
uniformly random key the expected null rate is $p=\gamma$ for $L\ge1$, but a
fixed key and finite language distribution can be offset. We therefore estimate
$p$ on a disjoint reference corpus and calibrate deployment thresholds
empirically; for $L{=}0$ this also corrects nonuniform natural byte frequencies.
The normalized $z$ is used as a ranking score, without assuming overlapping
byte windows are independent.
Detection never invokes the tokenizer; it is tokenization-invariant by
construction. $L{=}0$ is a context-free \emph{byte-Unigram}; $L\ge1$ is
context-byte-keyed. This yields two byte-level design points in the
(context $\times$ signal) space: context-free byte keying for $L{=}0$ and
context-dependent byte keying for $L\ge1$. KGW and Unigram provide the
corresponding token-level comparisons.
The default canonicalizer deletes whitespace and Unicode format characters. A
stress-test variant retains only Unicode alphabetic characters. It is exactly
invariant to punctuation, digit, emoji, whitespace, and format insertion, but
can erase semantically important content and is not suitable for code or
number-sensitive domains.

\section{Local scoring guarantees}\label{sec:theory}
\begin{definition}[Aligned attack range]
Suppose one canonical edit replaces $\byte(x)[a:a+r]$ by $s$ bytes. Pair the
unchanged prefix at equal indices and the unchanged suffix after shifting its
indices by $s-r$. $\AR(\delta)$ is the larger number of old or new scored
endpoints whose length-$(L+1)$ window has no identical aligned partner. Then
$|S(\tilde x)-S(x)|\le\AR(\delta)$.
\end{definition}

\begin{proposition}[$\Can$-invariance $\Rightarrow$ range $0$]\label{prop:zero}
If $\Can(\tilde x)=\Can(x)$ then $\byte(\tilde x)=\byte(x)$, so
$\green_t(\tilde x)=\green_t(x)$ for all $t$; hence $\AR=0$ and $z$ is unchanged,
for every $L$. In particular, whitespace/zero-width insertion has attack range
$0$ and leaves the detection statistic \emph{exactly} invariant.
\end{proposition}

\begin{proposition}[Locality of visible edits]\label{prop:local}
If an edit replaces $r$ consecutive canonical bytes by $s$ bytes, then
$\AR\le\max(r,s)+L$. Old windows ending in $[a,a+r+L)$ and new windows ending
in $[a,a+s+L)$ are the only unpaired windows; all later suffix windows are
identical after the index shift. A one-byte substitution, insertion, or deletion
therefore has range at most $1+L$. Contrast token-keyed schemes, where a
character edit can force nonlocal re-tokenization.
Moreover, the one-sided loss obeys
$S(\tilde x)\ge S(x)-(r+L)$: only old unpaired windows can remove existing
green indicators, whereas new unpaired windows can only add nonnegative terms.
\end{proposition}

\begin{corollary}[Detectability--robustness tradeoff]\label{cor:tradeoff}
Under $m$ localized edits with $r\le q$, the one-sided bound gives
$S(\tilde x)\ge S(x)-m(q+L)$, so the worst-case robustness bound weakens with
$L$. Empirically, $L{=}1$ improves detectability over $L{=}0$ on our tested
models and domain; that detectability ordering is an observation, not a theorem.
\end{corollary}

\begin{proposition}[Deterministic byte-edit radius]\label{prop:cert}
Let a document of canonical length $n$ have green count $S$ and count threshold
$h(n)=p(n-L)+\tau\sqrt{(n-L)p(1-p)}$, with $\tau\ge0$. If each elementary edit replaces at most
$r_{\max}$ bytes by at most $s_{\max}$ bytes, detection is guaranteed for every
$m$ satisfying
\[
S-m(r_{\max}+L)>
h(n+ms_{\max}).
\]
The deterministic radius is the largest such integer; $\Can$-invariant edits do not
count against it. If edits have an additional length-growth bound $g$, replace
$s_{\max}$ by $g$. Fixed-length edits have $g=0$ and recover the count margin
divided by the per-edit bound.
\end{proposition}
Proofs are given in Appendix~\ref{app:proofs}. The bound is unaffected by
tokenization and by unbounded
$\Can$-invariant edits, unlike token-level guarantees, and it shrinks in $L$,
restating Corollary~\ref{cor:tradeoff}.

\paragraph{Comparison with detection-side repair.} Normalization can undo
invariant perturbations for token-keyed schemes; exact recovery of the original
text is sufficient but not necessary to preserve detection. \tiwm{} instead
keeps the score unchanged under $\Can$-invariant edits
(Prop.~\ref{prop:zero}) and bounds the count loss from other local edits
(Prop.~\ref{prop:local}). These properties do not establish superiority over
defended token baselines. Visible insertions outside $D$ are not invariant and
must be evaluated separately.

\section{Experimental setup}
The primary experimental campaign, including generation, attacks, and evaluation,
ran entirely on a single Apple MacBook Pro with an M5 Pro chip and 64\,GB of
unified memory, without remote inference or external GPU compute. This hardware
description applies to the primary campaign, not all historical diagnostics
retained in the appendices.

We use Qwen3-8B and Gemma 4 E4B with 128 matched writing prompts and a
200-token generation limit. Both run locally with 4-bit checkpoints in MLX,
temperature 0.8 and top-$p$ 0.95. The primary comparison uses three predeclared
keys (15485863, 32452843, 49979687), KGW and Unigram at $\delta=1$, and
default and alphabetic \tiwm{} $L=1$ at $\delta=1.25$. Each key defines a
separate watermark, not an ensemble. We additionally sweep
$\delta\in\{0.5,1,1.5,2\}$ and generation limits 100/400 at the first key.
Prompts and sampling seeds are paired across watermarked schemes; unwatermarked
outputs use a separate fixed seed schedule and are shared across keys.

The white-box compound GA uses population 40, 30 generations, and a fixed
genome-to-text rendering. It chooses word positions and typo, deletion, swap,
zero-width insertion, or homoglyph operations to minimize the target score.
Deletion (DE) and Unicode normalization (UN) are applied inside baseline
fitness evaluations. Primary settings receive 2\%, 5\%, and 30\% word-edit
budgets; sweeps and length controls receive 5\%. Visible insertion is a
separate greedy punctuation/digit/emoji attack at 30\% of word boundaries.

AUC uses unwatermarked model outputs as negatives, unless explicitly stated
otherwise, and assigns half credit to ties. Primary table entries average
per-key AUCs; 95\% intervals resample positive and negative prompt indices
jointly across the fixed keys and compared methods (2{,}000 replicates).
They do not quantify uncertainty over a population of keys. PPL ratios divide
the mean document PPL by that of unwatermarked outputs from the same model;
primary tables then average those ratios over keys. These are automatic
distortion proxies, not human quality measurements.

The new locality and method audits reuse 10{,}000 human-text documents from
IMDb and English, French, and Chinese Wikipedia
\citep{maas2011learning,wikimedia2023wikipedia}, with a frozen source-stratified
5{,}000/5{,}000 calibration/evaluation split. Thresholds use a conservative
empirical quantile and strict exceedance; rates include exact binomial
intervals. Reusing this development corpus is not fresh confirmatory testing.
The earlier byte-detector audit uses a distinct 4{,}939/5{,}061 split and is
identified separately below. Earlier stochastic-fitness GA runs are superseded:
their winning genomes did not identify a unique scored text.

EXP-edit is the most relevant edit-robust baseline. We implement its published
inverse-transform sampler and edit-distance recurrence, match the common
sampling truncation, and calibrate every document against 64 alternate keys.
This algorithm-equivalent implementation uses a different pseudorandom number
generator from the authors' software, so it is not bitwise identical. On this
instruction setting, EXP-edit does not establish clean detectability; we report
that negative result rather than interpreting its attacked score as robustness.

\section{Results}\label{sec:results}

\begin{table}[H]\centering\small
\caption{Qwen3-8B, 128 prompts per key, three fixed keys, 200-token limit.
Entries average per-key statistics. Brackets give 95\% prompt-bootstrap
intervals conditional on these keys. PPL is a ratio to unwatermarked output;
settings are not matched on clean detectability.}\label{tab:main}
\begin{tabular}{lcccc}
\toprule
scheme & base-AUC & PPL & AUC, 5\% edits & AUC, 30\% edits\\
\midrule
\input{artifacts/paper_campaign_20261004/primary_qwen.tex}
\end{tabular}
\end{table}

\begin{figure}[H]\centering
\includegraphics[width=\linewidth]{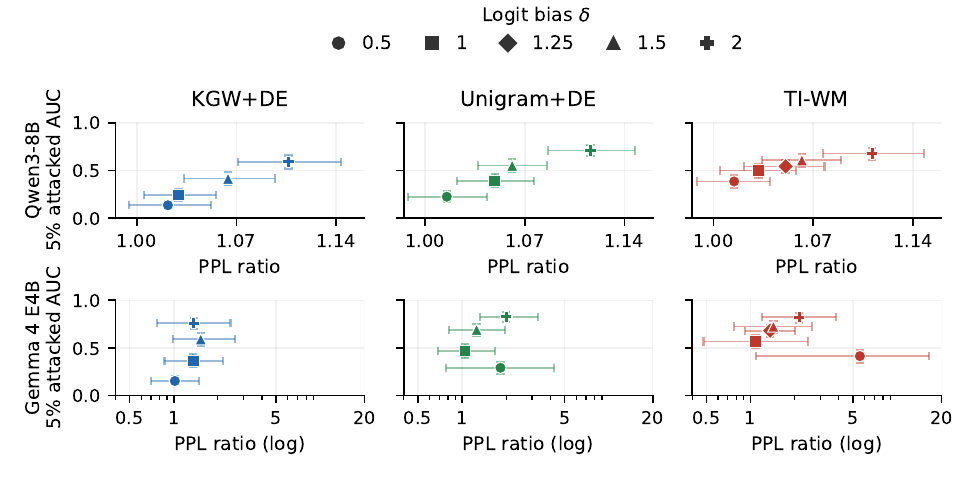}
\caption{First-key bias sweeps: 128 prompts per model, 5\% deterministic
attack. Shapes denote $\delta$; bars are pointwise 95\% bootstrap intervals.
PPL is relative to the unwatermarked mean. Axes are shared within rows;
Gemma's PPL axis is logarithmic. Equal PPL does not establish equal
detectability or perceptual quality.}\label{fig:evidence}
\end{figure}

\paragraph{Deterministic compound attack.} Table~\ref{tab:main} replaces the
earlier single-key headline. At 5\% edits, default \tiwm{} exceeds Unigram+DE
by \QwenGainFive{} (paired difference and 95\% interval). Its per-key AUC is
near chance, however, and higher ranking than an attacked baseline is not
evidence of useful detection power. At 30\%, all methods are weak. This attack
optimizes detector removal without a quality constraint.

At 2\%, the corresponding Qwen difference is \QwenGainTwo{}, which includes
zero. The new Gemma difference is \GemmaGainTwo{}, which does not. Thus even
the low-budget ordering depends on the model and tested operating point.

\begin{table}[H]\centering\small
\caption{Gemma 4 E4B replication with the same 128 prompts, three keys,
defenses, and uncertainty convention as Table~\ref{tab:main}. Mean PPL is
particularly sensitive to rare outputs in this model.}\label{tab:gemma}
\begin{tabular}{lcccc}
\toprule
scheme & base-AUC & PPL & AUC, 5\% edits & AUC, 30\% edits\\
\midrule
\input{artifacts/paper_campaign_20261004/primary_gemma.tex}
\end{tabular}
\end{table}

\paragraph{Model-family replication and distortion.} On Gemma, default
\tiwm{} exceeds Unigram+DE by \GemmaGainFive{} at 5\%. The mean per-key PPL
ratios are \GemmaBytePPL{} and \GemmaUnigramPPL{}, respectively, compared
with \QwenBytePPL{} and \QwenUnigramPPL{} on Qwen. These are not
quality-matched or detectability-matched results. Unigram+UN is slightly
stronger than Unigram+DE on Gemma, so DE is not a universally strongest defense.
The first-key sweeps in Figure~\ref{fig:evidence} show why a single operating
point is insufficient: stronger token biases recover much of the attacked
ranking, while Gemma's PPL curve is nonmonotone and has wide intervals.
Even the 5\% attacks increase same-model PPL substantially
(Appendix~\ref{app:campaign}); they are not validated as imperceptible edits.

\paragraph{EXP-edit boundary.} At matched sampling and a $+5.1\%$ PPL cost,
our EXP-edit implementation obtains clean AUC .474 $[.374,.577]$ on 64 Qwen
prompts after per-document alternate-key calibration. Its 5\%-attack AUC is
.098 $[.049,.155]$, but that number is not evidence of removal robustness
because the clean detector is already at chance. This result is consistent
with the reported difficulty of watermarking low-entropy instruction prompts
using this construction~\citep{kuditipudi2024robust} and leaves a controlled
edit-robust comparison open.

\paragraph{Locality controls.} We compare token, whitespace-delimited word,
code-point, and byte units with a shared keyed hash and generation rule
(Table~\ref{tab:locality}). The binary rule rewards a positive increment in
centered green count. All arms, including unwatermarked output, use the same
32-candidate base-logit shortlist, so these are controlled variants, not
reimplementations of the headline baselines. Candidate rewards use joint
prefix decoding rather than decoding tokens in isolation.

Character-scale controls retain more attacked ranking than token or word
controls on both models. This is consistent with locality, but not a monotone
token-to-word-to-character-to-byte law: byte scoring is not better than
code-point scoring. They coincide on ASCII under the shared hash; Unicode
characters and non-ASCII attack substitutions can separate them. Context order
one spans different amounts of text, word boundaries require separate handling,
and clean AUC differs across units. Thus the controls narrow, rather than fully
resolve, causal attribution.

\begin{table}[H]\centering\small
\setlength{\tabcolsep}{4pt}
\caption{Shared-rule locality controls, first key, 128 prompts, 200-token
limit. Attack is the 5\% deterministic GA; brackets give pointwise 95\%
document-bootstrap intervals. All binary variants use bias 1.}
\label{tab:locality}
\begin{tabular}{lcccc}
\toprule
& \multicolumn{2}{c}{Qwen} & \multicolumn{2}{c}{Gemma}\\
unit & clean AUC & attacked AUC & clean AUC & attacked AUC\\
\midrule
\input{artifacts/paper_campaign_20261004/locality.tex}
\end{tabular}
\end{table}

\paragraph{Scope of the controls.} The shared-rule controls use a different
hash and generation protocol from \tiwm{}. Their results are consistent with
locality, but word boundaries, context span, and clean detectability remain
confounded. Separate code-point and reward-shaping refinements do not establish
a calibrated-power improvement; their full results are in
Appendices~\ref{app:refinements} and \ref{app:rankpilot}.

\begin{center}\small
\refstepcounter{table}\label{tab:tpr}
\begin{minipage}{.96\linewidth}\centering
Table~\thetable: Earlier byte-detector null audit, 4{,}939 calibration and
5{,}061 evaluation documents, at nominal $10^{-3}$ FPR. Max LOO
is the worst leave-one-source-out FPR. Repeat FPR repeats a 128-character span
eight times. Alphabetic entries use the expanded canonicalizer.\par\smallskip
\begin{tabular}{lccc}
\toprule
detector & pooled FPR & max LOO FPR & repeat FPR\\
\midrule
default $L=1$ & .00158 & .0882 & .01976\\
context-conditioned $L=1$ & .00138 & .0223 & .02351\\
alphabetic $L=1$ & .00158 & .2838 & .01265\\
alphabetic, context-conditioned & .00079 & .0286 & .02035\\
unique-window $L=1$ & .00040 & .00477 & .00020\\
\bottomrule
\end{tabular}
\end{minipage}
\end{center}

\paragraph{Visible insertion (earlier first-key diagnostic).} At a 30\% word-boundary budget, inserting
punctuation, digits, or emoji reduces default \tiwm{} from base-AUC .732 to
.002. Alphabetic-only scoring is exactly invariant on all 128 outputs and stays
at .773. This fix is narrow: it may erase meaningful numbers and punctuation,
and it does not prevent the compound attacks in Tables~\ref{tab:main}--\ref{tab:gemma}.

\begin{table}[H]\centering\small
\caption{Adaptive visible-character insertion at a 30\% word-boundary budget
($n=128$, earlier first-key Qwen diagnostic, not averaged over keys).
Exact is the fraction whose score is unchanged.}\label{tab:visible}
\begin{tabular}{lccc}
\toprule
canonicalizer & base-AUC & attacked-AUC & exact\\
\midrule
default & .732 & .002 & .000\\
alphabetic-only & .773 & .773 & 1.000\\
\bottomrule
\end{tabular}
\end{table}

\paragraph{Operating points.} The original byte-detector audit in
Table~\ref{tab:tpr} limits the interpretation of the AUC gains: default $L=1$
has a worst held-out-source FPR of .0882 at nominal .001, and repetition also
raises false positives. Earlier default and alphabetic context-conditioned
detectors each detected none of 128 outputs at a human-calibrated 1\% threshold.
A pooled calibration rate near nominal therefore does not establish source
robustness. Separate bounded code-point designs also fail their clean-power
target (Appendix~\ref{app:refinements}); that result is not a measurement of the
original \tiwm{} detector. The experiments do not establish deployable low-FPR
detection for the proposed construction.

\paragraph{Paraphrase (orthogonal threat).} Locality does not protect against
rewriting that changes most scored windows. Our earlier paraphrase diagnostics
do not establish an advantage for \tiwm{}, and the present campaign does not
evaluate a quality-preserving paraphraser. Semantic watermarking addresses a
different threat; general limitations on robustness to rewriting further
constrain what locality alone can establish~\citep{zhang2024watermarks}.

\paragraph{Spoofing and a mitigation.}\label{sec:spoof}
Green-list watermarks are known to be forgeable~\citep{jovanovic2024watermark}.
Our earlier diagnostic estimates byte greenness from watermarked text and
case-swaps natural text toward estimated-green bytes. We mitigate this channel by
\emph{case-folding} $\Can$ (lowercase before keying), so a case-swap forge is a
no-op on the canonical form. Thus, case-folding mitigates this demonstrated spoof,
but the result does not establish resistance to a different adaptive forge.
A fully adaptive forger must find a different cheap channel surviving
case-folding; private/keyed detection is an orthogonal further defense.

\section{Discussion and limitations}
\paragraph{Comparison with token-level designs.}
The expanded experiment supports a narrow claim. At the primary 5\% operating
points, byte scoring retains more ranking signal than KGW or Unigram on both
models. Its PPL cost is comparable on Qwen but substantially larger and more
variable on Gemma. The 2\% difference is unresolved on Qwen, not on Gemma.
No comparison is matched simultaneously on base AUC and PPL. EXP-edit
fails to establish clean detectability in this setting, so it does not provide
a usable edit-robust comparison. We therefore report a bounded byte-locality
effect, not general superiority over edit-robust watermarking.

\paragraph{Relation to semantic and repair approaches.}
The paraphrase experiment does not show an advantage for \tiwm{}; semantic
watermarks such as SIR and PASA address that separate threat and may be
composable with canonicalized byte scoring. Tokenization rollback instead tries
to reconstruct the original token sequence after an edit. \tiwm{} avoids that
inversion step for $\Can$-invariant edits and bounds the effect of other local
edits directly. These approaches therefore address different failure modes.

\paragraph{Method refinement and remaining bottleneck.}
The proposed method separates tokenizer dependence from byte-local scoring;
reliable calibration and generation efficacy remain additional requirements.
Neither the separate reward-shaping pilot (Appendix~\ref{app:rankpilot}) nor
the bounded code-point screen (Appendix~\ref{app:refinements}) resolves them.
Their failures do not show that the original construction cannot be improved.
Uniform hash
colors over random keys do not imply a language-independent text null for one
fixed key. Repetition caps change both effective signal and variance, while
alphabetic filtering exchanges an insertion channel for a restricted text
representation. These results argue for testing calibration and generation
efficacy jointly, not treating either as a consequence of smaller signal units.

\paragraph{Limitations.}
\tiwm{} does not claim universal robustness: quality-preserving paraphrase can
challenge non-semantic watermarking~\citep{zhang2024watermarks}. Our method
contribution is canonical-byte keying and tokenizer-free detection with local
scoring guarantees, not a proof of useful detection after every edit. The headline
uses the same 128 prompts across two models and three predeclared keys, not
independent prompt samples for every key. Bias sweeps and length checks use
one key. Shared-rule controls restrict candidate sampling and are not a causal
test of signal granularity at matched power. Human-text audits reuse development
data; a fresh held-out source and a larger quality-constrained attacker remain
necessary. The twelve-variant screen tests specific rewards and biases, not all
possible code-point methods.
No experiment resolves operational FPRs near $10^{-5}$.

The deletion set $D$ is an unavoidable attack surface. A character outside it
changes the score; expanding it can erase meaning. Alphabetic-only scoring
handles the tested insertion attack but is inappropriate for code, mathematics,
and number-sensitive text, and worsens held-out-language calibration. The
human-null audit shows that reference calibration does not solve domain shift
or repetition. At calibrated thresholds, detection power is inadequate.

The 30\% GA is a detector stress test, not an imperceptible attack. A historical
quality-gated AUC table was withdrawn because cached fitness values differed
from scores of its saved texts. Two model judges never preferred those attacked
texts in an order-consistent comparison
(Appendix~\ref{app:local}). No human evaluation is included. Spoofing remains a
shared green-list vulnerability; case-folding closes only the tested case-swap
channel. Paraphrase and spoofing results are limited diagnostics; they do not
establish robustness to either threat. Distillation is not evaluated.

\section{Conclusion}
We introduced \tiwm{}, a watermarking method that embeds signal during token
generation and detects it through canonical-byte keying and scoring. Its
construction gives exact invariance to canonicalizer-preserving edits and
bounds the count loss from other localized canonical-byte edits. On two models
and three fixed keys, it retains more ranking signal than the tested token
baselines at the primary 5\% attack settings. Bias sweeps and adaptive stress
tests delimit that gain: distortion is model dependent, visible insertions can
bypass the default canonicalizer, and source calibration and low-FPR power
remain unresolved. \tiwm{} provides a concrete route to tokenizer-independent
local scoring; turning those properties into practical detection requires
further advances in signal generation, quality-preserving robustness, and
calibration.

\appendix
\section{Proofs}\label{app:proofs}

\paragraph{Proposition~\ref{prop:zero}.}
$\Can(\tilde x)=\Can(x)$ implies equal UTF-8 byte strings. Every scored byte and
its preceding $L$ bytes are therefore identical, so every indicator, the green
count, and the normalized score are unchanged.

\paragraph{Proposition~\ref{prop:local}.}
Suppose $r$ old bytes beginning at $a$ are replaced by $s$ new bytes. In the old
string, only windows ending in $[a,a+r+L)$ can touch the replaced interval or
its following context, giving at most $r+L$ unpaired windows. The analogous new
interval gives at most $s+L$. Earlier windows are identical at equal indices;
later windows consist entirely of the common suffix and are identical after an
index shift of $s-r$. Hence the aligned range is at most
$\max(r+L,s+L)=\max(r,s)+L$. Since paired indicators cancel and each unpaired
sum lies between zero and its number of terms, the absolute green-count change
is bounded by the same quantity.

\paragraph{Proposition~\ref{prop:cert}.}
Apply the one-sided part of Proposition~\ref{prop:local} sequentially:
after $m$ edits, the green count is at least
$S-m(r_{\max}+L)$. The canonical length is at most
$n+ms_{\max}$, since the removal bound permits removing zero bytes.
For $\tau\ge0$, the count threshold $h(n)$ is
nondecreasing in $n$. Comparing the lower green-count bound with the threshold
at the maximum possible length proves the stated bound. A
$\Can$-invariant edit changes neither quantity and therefore consumes no radius.

\section{Expanded campaign details}\label{app:campaign}

\paragraph{Length and numerical radii.}
Table~\ref{tab:length} varies the generation limit, not a forced output length.
Early stopping makes the observed mean smaller than the limit. Both models
show improved attacked ranking with longer outputs at the first key. The
all-document median substitution radius becomes nonzero at the 400-token
limit, but the radius is evaluated at a fixed $z>4$, not a calibrated human
false-positive threshold. It covers one-byte substitutions in the canonical
string, not arbitrary word edits or multi-byte character replacements. Zero
entries include documents that were not detected before any edit; detected
counts are therefore reported separately.

\begin{table}[H]\centering\footnotesize
\setlength{\tabcolsep}{3pt}
\caption{First-key length scaling ($n=128$ per row). Observed tokens and radii
refer to default \tiwm{} at $\delta=1.25$. Both attacked AUC columns use 5\%
word edits; Unigram uses DE and $\delta=1$. Radius is the median over all
documents under the fixed-threshold canonical-substitution bound.}
\label{tab:length}
\begin{tabular}{lrrcrcr}
\toprule
model & limit & mean tokens & \tiwm{} AUC [95\% CI] & Unigram & $z>4$ & radius\\
\midrule
\input{artifacts/paper_campaign_20261004/length.tex}
\end{tabular}
\end{table}

\paragraph{Quality and sampling controls.}
At the first key and primary biases, the 5\%-attack/original mean PPL ratio
for default \tiwm{} is 1.833 on Qwen and 2.397 on Gemma; Unigram+DE gives
1.887 and 2.514. A small edit budget is therefore not a demonstrated quality
constraint. PPL is evaluated on generated continuations without the prompt.
The shared-rule controls use a different sampling protocol from the
full-vocabulary primary comparison: candidates are restricted to the 32
largest base logits before watermark reward, then sampled at the same
temperature and top-$p$. A wider, 128-candidate check uses 32 prompts per
model. Its broad attack ordering persists, but its smaller sample does not
establish equivalence to unrestricted generation.

\paragraph{Gemma PPL tail audit.}
\input{artifacts/paper_campaign_20261004/gemma_ppl_audit.tex}
Inspection of the highest-PPL saved texts finds stray suffixes and repeated
punctuation in some outputs, but also ordinary-looking prose. The scores are
completion-only, not prompt-conditioned; high PPL alone does not establish
poor human-perceived quality. No documents are excluded from the primary
results. This saved-score audit explains the unstable arithmetic mean, not
why a particular generation received a high model score.

\paragraph{Control statistic.}
For each distinct context-and-unit window $w$, let $m_w$ be its multiplicity
(capped where specified), and let $g_k(w)$ be its keyed BLAKE2 green indicator.
The shared-rule score is
\[
\frac{\sum_w m_w(g_k(w)-\gamma)}
{\sqrt{\gamma(1-\gamma)\sum_w m_w^2}},\qquad \gamma=\tfrac12.
\]
Byte and code-point units share an integer encoding. Token units are obtained
by tokenizing canonicalized text; word units are split before whitespace
deletion to avoid merging the document. Thus the token control is not KGW,
and the byte control is not the original SHA-based \tiwm{} detector.
Default control filtering removes Unicode whitespace and format characters;
the historical byte implementation uses an explicit deletion list. These
distinctions prevent treating the control table as an exact decomposition of
the headline gain. The denominator describes variation over random hash
colors, not a guarantee over text sources for a fixed key.

\paragraph{Audit boundaries.}
Calibration documents are split within each source before thresholds are
computed. Leave-one-source-out thresholds omit that source from calibration
and are evaluated on its held-out documents. The pooled split has 5{,}000
evaluation documents, with 1{,}250 per source; such samples cannot resolve
operational guarantees near $10^{-5}$. The same corpus informed earlier
development, so these are diagnostic, not confirmatory, operating points.

\section{Follow-up designs, distinct from \tiwm{}}\label{app:refinements}
The following experiments alter the scored unit, generation reward, or detector.
They are not the original SHA-based byte-keyed \tiwm{} construction. Their
failures constrain these tested refinements, not every possible instantiation
of \tiwm{} or code-point watermarking.

\paragraph{Bounded code-point variants.} We test the candidate bias
\[
\delta\,\operatorname{clip}\!\left(
\frac{\Delta S-\gamma\Delta T}{\sqrt{\max(1,n)}},-2,2\right),
\]
with $\delta=1$, where $n$ counts added or removed window occurrences.
The detector divides the centered count by a random-key variance scale that
accounts for window multiplicity; it is not a fixed-key calibrated $p$-value.
We cross three keys with default filtering, letters-only filtering, and
letters-only filtering with window-count caps 2 or 4. Each window contributes
at most the cap across the whole document, in both generation and detection.
All twelve variants detect $0/128$ outputs at the development-calibrated
$10^{-3}$ threshold (each exact 95\% binomial interval is $[0,0.0284]$).
Letters-only variants preserve scores under the tested visible insertion but
still fail source-held-out calibration (Table~\ref{tab:methodscreen}).
This rejects these tested settings as a solution to calibrated power, not
code-point watermarking in general.

\begin{table}[H]\centering\footnotesize
\setlength{\tabcolsep}{4pt}
\caption{Bounded code-point screen on Qwen, 128 prompts per variant, three
keys. Ranges are observed minima/maxima across keys, not confidence intervals.
TPR is clean detection at a human-calibrated nominal $10^{-3}$ FPR; the count
is zero separately for every key. Max LOO is each key's worst held-out-source
FPR. Visible drop is clean minus insertion-attacked AUC.}
\label{tab:methodscreen}
\begin{tabular}{llcccc}
\toprule
filter & cap & 5\% AUC & detected & max LOO FPR & visible drop\\
\midrule
\input{artifacts/paper_campaign_20261004/method.tex}
\end{tabular}
\end{table}

\paragraph{Earlier generation-rule ablations.} We do not interpret $L=0$
versus $L=1$ as a clean frontier. Earlier short-text pilots found weak signal
under the $L=0$ majority rule, little benefit from natural-frequency balancing,
and repetition or PPL failure under an unbounded count reward. These motivate
the bounded screen above rather than establishing a general impossibility.

\paragraph{Follow-up reward-shaping pilot.} A separate hybrid multi-key pilot
rewards candidate bytes by their primary-minus-decoy detector-score increment.
Across three key seeds, it improves 5\%-attack AUC against held-out human text
relative to raw-count rewards, but calibrated TPR remains at most $2/64$ and is
zero for two seeds. This is a relative-ranking gain, not useful operating-point
power; details are in Appendix~\ref{app:rankpilot}.

\section{Earlier local diagnostics}\label{app:local}
These experiments use preserved same-model Qwen3-8B outputs and test mechanisms
in a resource-bounded setting. They precede the expanded campaign and are not
additional independent replications of its headline estimates. Sample sizes,
thresholds, and negative classes differ and are identified below.

\subsection{Detector-aligned reward pilot}\label{app:rankpilot}
This follow-up compares raw green-count rewards with an additive
primary-minus-decoy reward for a hybrid multi-key detector. Each byte reward is
the primary group's score increment minus the mean decoy increment, scaled by
the decoy increment standard deviation; candidate-token rewards sum these
increments across bytes. A per-step KL controller is held fixed. This reward is
a local approximation to the document-level studentized detector, not its exact
gradient. We test one 64-prompt run and two 32-prompt key-seed replications on
Qwen3-8B. For each seed, thresholds use a disjoint human-text calibration split
and are evaluated on held-out human text. Attacked AUC uses a deterministic
5\% white-box GA. Paired 95\% intervals resample prompts and negative documents.

\begin{table}[H]\centering\footnotesize
\setlength{\tabcolsep}{3pt}
\caption{Rank-aligned reward pilot. Clean AUC uses matched unwatermarked model
outputs; attacked AUC uses held-out human text. Each paired entry is raw/rank.
The gain column gives paired 95\% intervals for the rank-minus-raw difference.
TPR is measured after the attack; PPL is relative to unwatermarked text.}
\begin{tabular}{lccccc}
\toprule
seed ($n$) & clean AUC & 5\% human AUC & gain [95\% CI] & TPR & PPL ratio\\
\midrule
003 (64) & .896/.944 & .722/.797 & .075 [.050,.101] & 1/64, 2/64 & 1.258/1.265\\
004 (32) & .880/.962 & .305/.452 & .147 [.073,.223] & 0/32, 0/32 & 1.221/1.174\\
005 (32) & .832/.899 & .422/.599 & .177 [.123,.235] & 0/32, 0/32 & 1.201/1.164\\
\bottomrule
\end{tabular}
\end{table}

Clean AUC against held-out human text for raw/rank rewards is .799/.862,
.431/.586, and .585/.726 across the three seeds. The positive paired AUC
differences show a consistent relative benefit from reward alignment, but the
absolute human separation and threshold TPR remain inadequate and key
dependent. No human preference
evaluation was performed. These results motivate further method work; they do
not revise the headline claims or establish a deployable detector.

\subsection{Representative attack examples}\label{app:examples}
Table~\ref{tab:examples} places excerpts from the first prompt of three
completed campaign conditions side by side. Bracketed Unicode labels are
display substitutions, not literal text. The excerpt starts near the first
word-level difference; scores use the entire document. These examples do not
estimate attack frequency or text quality.

\begin{table}[H]\centering\footnotesize
\setlength{\tabcolsep}{4pt}
\caption{Saved Qwen campaign examples at the first key and 200-token limit.
The first two rows use default \tiwm{}; the last uses the bounded code-point
variant with letters-only filtering and no repetition cap. Its unchanged
score demonstrates invariance, not successful detection.}
\label{tab:examples}
\begin{tabular}{p{.14\linewidth}p{.385\linewidth}p{.385\linewidth}}
\toprule
setting & before & after attack \\
\midrule
\input{artifacts/paper_campaign_20261004/examples.tex}
\end{tabular}
\end{table}

\paragraph{Withdrawn quality-gated AUC table.}
The earlier candidate-pool selector used cached GA fitness values that do not
match scores recomputed on its saved attacked strings. We therefore withdraw
its AUC table and paired robustness claims rather than treating post-hoc
selection as a correctly optimized, quality-constrained attack. The judge
comparisons below concern the actual selected strings only; neither their
preferences nor PPL caps validate that withdrawn frontier.

\begin{table}[H]\centering\small
\caption{Historical cross-family quality diagnostic on saved, quality-gated pairs
using \texttt{Qwen/Qwen3.8-27B} and
\texttt{google/gemma-4-E4B-it}~\citep{qwen38model,gemma4model} (128
scheme--document pairs per gate). Each pair is judged in both A/B orders;
only decoded order-consistent pairs are counted. The selector retained the exact
original in 110, 31, and 1 pairs at the three gates. Neither judge replaces
human evaluation.}
\begin{tabular}{llrrr}
\toprule
judge & PPL cap & consistent & attack/original/tie & position gap\\
\midrule
Qwen3.8-27B & $1.25\times$ & 122 & 0 / 10 / 112 & .000\\
             & $1.5\times$  & 100 & 0 / 60 / 40  & .008\\
             & $2\times$    & 103 & 0 / 95 / 8   & .032\\
Gemma 4 E4B  & $1.25\times$ & 125 & 0 / 0 / 125  & .000\\
             & $1.5\times$  & 106 & 0 / 3 / 103  & .008\\
             & $2\times$    & 84  & 0 / 6 / 78   & .008\\
\bottomrule
\end{tabular}
\par\smallskip
Qwen pooled order consistency is 0.953, 0.781, and 0.805; Gemma obtains 0.977,
0.828, and 0.656. Both produce zero order-consistent attack wins, but their
different original-versus-tie rates expose model-judge sensitivity. The strict
gate also mostly prevents an eligible change. These are negative diagnostics:
they do not validate perceptual equivalence.
\end{table}

\begin{table}[H]\centering\small
\caption{Local quality diagnostics on 32 prompts. PPL is scored by the same
Qwen3-8B model; similarities compare the saved attacked text with its source.
These automatic metrics motivate, but do not replace, human evaluation.}
\begin{tabular}{lccc}
\toprule
setting & mean PPL & char. similarity & word similarity\\
\midrule
unwatermarked & 6.00 & -- & --\\
KGW / Unigram & 6.35 / 6.29 & -- & --\\
\tiwm{} $L=0$ / $L=1$ & 6.59 / 6.18 & -- & --\\
10\%-edit GA (scheme range) & 15.88--20.29 & 0.987--0.989 & 0.902--0.906\\
30\%-edit GA (scheme range) & 36.24--50.20 & 0.965--0.970 & 0.717--0.736\\
\bottomrule
\end{tabular}
\end{table}

\begin{table}[H]\centering\small
\caption{Matched-seed Qwen3-8B canonicalizer ablation on 32 64-token outputs.
These are shorter generations than Table~\ref{tab:main}, which explains the
different default $L=1$ base and attacked AUCs. The
deterministic compound GA uses a 30\% word-edit budget, population 40, and 30
generations. Default removes whitespace/format characters; UN-style also
reverses the attack's confusable map.}
\begin{tabular}{lcc}
\toprule
configuration & base-AUC & attacked-AUC\\
\midrule
\input{artifacts/paper_campaign_20261004/canonicalizer.tex}
\end{tabular}
\par\smallskip
The small, non-monotone differences do not attribute compound-attack
robustness to canonicalization. Canonicalization gives exact protection only
for the operations it removes.
\end{table}

At the fixed $z>4$ threshold, the deterministic one-byte-substitution radius
is vacuous for all 32 local $L=0$ outputs because none is initially detected.
Only one of 32 $L=1$ outputs is detected, and its radius is one. These short-text
results show that the bound is a worst-case accounting result, not evidence of
a practically large robustness radius.

Using token-sequence Levenshtein alignment, one zero-width insertion changes a
mean of $3.07$ Qwen token IDs (median $3$, maximum $5$) on 100 preserved local
texts. The earlier positional mismatch diagnostic gave $26.24$ because it
counted the shifted suffix; we do not use that number as evidence. The same
insertion leaves \tiwm{}'s canonical byte score exactly unchanged.

\bibliographystyle{plainnat}
\bibliography{references}
\end{document}

%% file: artifacts/paper_campaign_20261004/numbers.tex
\newcommand{\QwenByteAUC}{0.530}
\newcommand{\QwenBytePPL}{1.050}
\newcommand{\QwenUnigramAUC}{0.410}
\newcommand{\QwenUnigramPPL}{1.056}
\newcommand{\QwenGainTwo}{0.028 [-0.025,0.080]}
\newcommand{\QwenGainFive}{0.119 [0.062,0.173]}
\newcommand{\GemmaByteAUC}{0.664}
\newcommand{\GemmaBytePPL}{2.340}
\newcommand{\GemmaUnigramAUC}{0.489}
\newcommand{\GemmaUnigramPPL}{1.192}
\newcommand{\GemmaGainTwo}{0.069 [0.023,0.113]}
\newcommand{\GemmaGainFive}{0.175 [0.124,0.225]}

%% file: artifacts/paper_campaign_20261004/primary_qwen.tex
KGW & 0.857 & 1.049 & 0.237 [0.204,0.273] & 0.000 \\
KGW+DE & 0.850 & 1.049 & 0.259 [0.224,0.296] & 0.000 \\
KGW+UN & 0.857 & 1.049 & 0.274 [0.238,0.312] & 0.000 \\
Unigram & 0.799 & 1.056 & 0.374 [0.332,0.417] & 0.001 \\
Unigram+DE & 0.799 & 1.056 & 0.410 [0.366,0.454] & 0.003 \\
Unigram+UN & 0.799 & 1.056 & 0.409 [0.366,0.453] & 0.007 \\
\tiwm{} default & 0.754 & 1.050 & 0.530 [0.490,0.567] & 0.080 \\
\tiwm{} alphabetic & 0.762 & 1.059 & 0.513 [0.472,0.551] & 0.062 \\
\bottomrule

%% file: artifacts/paper_campaign_20261004/primary_gemma.tex
KGW & 0.894 & 1.696 & 0.336 [0.297,0.376] & 0.000 \\
KGW+DE & 0.892 & 1.696 & 0.362 [0.322,0.402] & 0.000 \\
KGW+UN & 0.894 & 1.696 & 0.373 [0.334,0.414] & 0.001 \\
Unigram & 0.857 & 1.192 & 0.461 [0.422,0.502] & 0.005 \\
Unigram+DE & 0.856 & 1.192 & 0.489 [0.450,0.531] & 0.006 \\
Unigram+UN & 0.857 & 1.192 & 0.495 [0.455,0.537] & 0.011 \\
\tiwm{} default & 0.848 & 2.340 & 0.664 [0.629,0.697] & 0.113 \\
\tiwm{} alphabetic & 0.844 & 2.124 & 0.635 [0.596,0.671] & 0.085 \\
\bottomrule

%% file: artifacts/paper_campaign_20261004/locality.tex
Token & 0.809 & 0.205 [0.153,0.264] & 0.886 & 0.331 [0.266,0.402] \\
Word & 0.793 & 0.214 [0.160,0.268] & 0.886 & 0.308 [0.245,0.369] \\
Code point & 0.721 & 0.493 [0.421,0.569] & 0.837 & 0.641 [0.573,0.708] \\
Byte & 0.721 & 0.467 [0.394,0.542] & 0.831 & 0.606 [0.539,0.673] \\
Code point, bounded & 0.764 & 0.525 [0.453,0.597] & 0.876 & 0.689 [0.624,0.751] \\
\bottomrule

%% file: artifacts/paper_campaign_20261004/length.tex
Qwen & 100 & 99.6 & 0.496 [0.422,0.564] & 0.370 & 10/128 & 0 \\
Qwen & 200 & 189.9 & 0.544 [0.472,0.612] & 0.392 & 40/128 & 0 \\
Qwen & 400 & 336.0 & 0.597 [0.529,0.662] & 0.469 & 80/128 & 6 \\
Gemma & 100 & 99.5 & 0.593 [0.519,0.659] & 0.428 & 9/128 & 0 \\
Gemma & 200 & 195.3 & 0.680 [0.613,0.743] & 0.466 & 45/128 & 0 \\
Gemma & 400 & 380.1 & 0.743 [0.679,0.800] & 0.549 & 102/128 & 12 \\
\bottomrule

%% file: artifacts/paper_campaign_20261004/gemma_ppl_audit.tex
At the first key, Gemma's arithmetic-mean PPL ratio for \tiwm{} is 5.57 at $\delta=0.5$, versus 1.08 at $\delta=1$. In the former setting, the largest document PPL accounts for 64.9\% of the PPL sum, and the five largest for 87.1\%. The geometric-mean PPL ratio is 1.10; this is a sensitivity diagnostic, not a replacement for the reported arithmetic-mean ratio.

%% file: artifacts/paper_campaign_20261004/method.tex
default & none & 0.525--0.579 & 0/128 & 0.0088--0.0664 & 0.264--0.381 \\
letters & none & 0.522--0.578 & 0/128 & 0.0056--0.0744 & 0.000--0.000 \\
letters & 2 & 0.351--0.374 & 0/128 & 0.0088--0.0552 & 0.000--0.000 \\
letters & 4 & 0.384--0.490 & 0/128 & 0.0088--0.0560 & 0.000--0.000 \\
\bottomrule

%% file: artifacts/paper_campaign_20261004/examples.tex
5\% compound & ...\$50 million in proposed transit funding, with proposals ranging from... & ...\$50 million in propo[U+0455]ed transit funding, with proposals ranging from... \\
\multicolumn{3}{p{.94\linewidth}}{Full-document score: $1.521 \to 0.464$.}\\[3pt]
30\% compound & ...City council members are currently debating how to allocate... & ...City council mrmbers aer currently debating hwo to allocate... \\
\multicolumn{3}{p{.94\linewidth}}{Full-document score: $1.521 \to -1.893$.}\\[3pt]
Visible insertion & ...City officials are currently locked in debate... & ...City. officials are. currently. locked in debate.... \\
\multicolumn{3}{p{.94\linewidth}}{Full-document score: $1.325 \to 1.325$.}\\[3pt]
\bottomrule

%% file: artifacts/paper_campaign_20261004/canonicalizer.tex
identity, $L=0$ & 0.615 [0.470,0.758] & 0.213 [0.110,0.338] \\
default, $L=0$ & 0.641 [0.503,0.772] & 0.242 [0.136,0.367] \\
UN-style, $L=0$ & 0.641 [0.503,0.772] & 0.257 [0.148,0.384] \\
identity, $L=1$ & 0.672 [0.527,0.803] & 0.163 [0.067,0.269] \\
default, $L=1$ & 0.708 [0.566,0.832] & 0.078 [0.022,0.162] \\
UN-style, $L=1$ & 0.708 [0.566,0.832] & 0.113 [0.039,0.215] \\
\bottomrule